\documentclass[12pt]{article} % A4 paper and 11pt font size

\usepackage[T1]{fontenc} % Use 8-bit encoding that has 256 glyphs
\usepackage{fourier} % Use the Adobe Utopia font for the document - comment this line to return to the LaTeX default
\usepackage[english]{babel} % English language/hyphenation
\usepackage{amsmath,amsfonts,amsthm} % Math packages
\usepackage{graphicx, floatrow, subfig}
\usepackage{lipsum} % Used for inserting dummy 'Lorem ipsum' text into the template
\usepackage{natbib}
\usepackage{graphicx}
\usepackage{xcolor}
\usepackage{sectsty} % Allows customizing section commands
\allsectionsfont{\centering \normalfont\scshape} % Make all sections centered, the default font and small caps
\usepackage{hyperref}
\usepackage{fancyhdr} % Custom headers and footers
\numberwithin{equation}{section} % Number equations within sections (i.e. 1.1, 1.2, 2.1, 2.2 instead of 1, 2, 3, 4)
\numberwithin{figure}{section} % Number figures within sections (i.e. 1.1, 1.2, 2.1, 2.2 instead of 1, 2, 3, 4)
\numberwithin{table}{section} % Number tables within sections (i.e. 1.1, 1.2, 2.1, 2.2 instead of 1, 2, 3, 4)

\usepackage[normalem]{ulem}

\newcommand{\horrule}[1]{\rule{\linewidth}{#1}} % Create horizontal rule command with 1 argument of height

\title{	
\normalfont \normalsize
\horrule{0.5pt} \\[0.4cm] % Thin top horizontal rule
\huge Are Nanoflares Responsible for Coronal Heating? 
\\ % The assignment title
\horrule{2pt} \\[0.5cm] % Thick bottom horizontal rule
}

\author{Loukas Vlahos$^1$, Heinz Isliker$^1$ and Nikos Sioulas$^2$ \\$^1$Departmet of Physics, Aristotle University\\54124 Thessaloniki, Greece\\ $^2$ Earth, Planetary, and Space Sciences, \\
University of California, Los Angeles \\
Los Angeles, CA 90095, USA} % Your name

\date{\normalsize\today} % Today's date or a custom date

\begin{document}

\maketitle % Print the title

%----------------------------------------------------------------------------------------
%	PROBLEM 1
%----------------------------------------------------------------------------------------

	\cite{Parker83} suggested a mechanism for the formation of current sheets (CSs) in the solar atmosphere. His main idea was that tangling of coronal  magnetic field lines by  photospheric  random flows facilitates the continuous formation of CSs in the solar atmosphere. This part of his idea represents one of the many ways by which the turbulent convection zone drives the formation of coherent structures and CSs in the solar atmosphere. Other mechanisms include emerging magnetic flux, interaction of current filaments, and explosive magnetic structures. However, there are two unproven assumptions in the initial idea of Parker for the coronal heating through nanoflares that have to be re-examined. They are related to his suggestion that {\bf ALL} CSs formed are led to magnetic reconnection and that magnetic reconnection heats the plasma in the solar atmosphere. Let us discuss these  two assumptions briefly in this  short comment:
	
	\begin{enumerate}
		\item Are all coherent structures and CSs formed by the turbulent convection zone reconnecting~?  Do  non-reconnecting CSs play a role in the heating of the corona~? 
		\item Does magnetic reconnection heat the plasma~? 
	\end{enumerate}

	We claimed in a recent article \citep{Vlahos21} that in a turbulent plasma only a fraction (if any) of the CSs or current filaments formed by the turbulent convection zone, as proposed by Parker, are going to reconnect. The non-reconnecting CSs in the turbulent part of the solar atmosphere play the dominant role in coronal heating. \cite{Einaudi21} have recently analyzed the formation of non-reconnecting CSs  (referring to them as "elementary events" instead of nanoflares). They have performed a careful examination  of the coherent structures formed inside the magnetic topology  used for their analysis and suggest that the resistive heating ($Q_i=\eta j^2 $, where $\eta$ is the resistivity and $j$ the current) due to the filamentation of the current and the formation of non-reconnecting CSs is sufficient to heat the corona.  Fig.\ 7 in their article presents the density distribution of the "elementary events" (non-reconnecting CSs). The question of how the associated distribution of resistive electric fields ($E_{eff}=\eta j$), randomly appearing in non-reconnecting CSs and interacting stochastically with particles, heat the coronal plasma, was not analyzed in the study of \cite{Einaudi21}. 
		
		\vspace{0.2cm}
		
		Let us now discuss the second question. 	
		
		\vspace{0.2cm}
		
		Recent numerical simulations have confirmed the idea that an isolated reconnecting CS systematically accelerates particles  (first-order Fermi acceleration), see Figure 3b in \cite{LiXiaocan2019} and Figure 3b in \cite{ Arnold21}, here presented below in Fig.\ \ref{f2}.

     \begin{figure}[ht!]\includegraphics[width=0.8\textwidth]{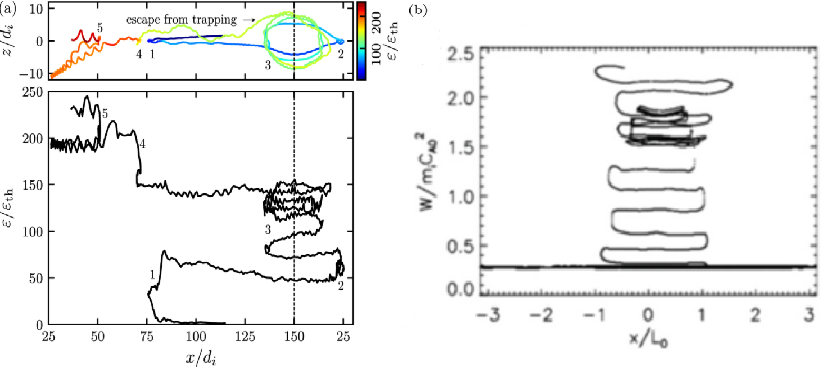}
 	\caption {Typical examples of particle energization in reconnecting current sheets: Systematic acceleration of particles in isolated and periodic Particle In Cell simulations: (a) Fig. 3 from \cite{LiXiaocan2019}, (b) Fig. 3b form \cite{Arnold21}.  } \label{f2}
      \end{figure} 
         
         \vspace{0.5cm}

	Both studies indicate that isolated reconnecting CSs result in a steady-state kinetic energy distribution that exhibits an extended power-law tail. However, Maxwellian fits to the steady state distribution at low energies indicate that isolated reconnecting CSs cannot heat the plasma on the kinetic level. On the other hand, in a 3D turbulent reconnection environment, where reconnecting CSs, non-reconnecting CSs, and coherent structures (turbulence) co-exist \citep{Comisso_2019, Isliker19}, one observes both, the heating, as well as the acceleration of particles. Consequently, as one moves from small (almost point-like) and isolated reconnecting CS, to 3D large-scale structures filled with reconnecting and non-reconnecting CSs, driven by the turbulent convection zone, something important is missing, which plays a crucial role in the impulsive heating and the shape of the steady-state distribution.
	 	
\begin{figure}[ht!]
\includegraphics[width=0.8\textwidth]{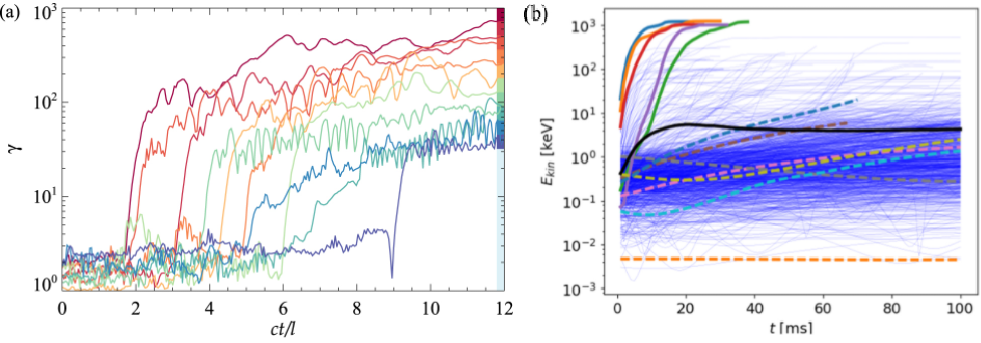}
        \caption{Examples of particle dynamics inside a turbulent reconnection environment from 3D numerical simulations. (a) Fig.\ 9a in \cite{Comisso_2019}: The systematic acceleration of the high-energy particles as they cross the reconnecting CSs is the main acceleration mechanism, accompanied by the stochastic interaction with the non-reconnecting CSs. In this study, the authors choose to discuss the role of the non-reconnecting CSs for high-energy particles, and they find that the stochastic interaction is a minor effect for the particles already accelerated by the reconnecting CSs. (b) Fig.\ 10 in \cite{Isliker19} (and see also Fig.\ 16 therein): Here we also observe systematic acceleration of the high energy particles crossing the reconnecting CSs. The high-energy particles escape from the acceleration volume soon after they have encountered the reconnecting CSs. The low energy particles interact with the non-reconnecting CSs and are impulsively heated, as shown in Fig.\ 11 in the same article.}\label{f1}
     \end{figure}

	 	 In \cite{Comisso_2019}, the turbulent reconnection environment was generated by perturbed magnetic fluctuations in a periodic, large simulation box.  \cite{Isliker19} use 3D MHD simulations to study the fragmentation of a large-scale CS generated by emerging magnetic flux and follow test particles inside an open large-scale 3D simulation volume around the fragmented CSs. The results from the PIC simulations and the MHD and test particle simulations are presented in the figure \ref{f1}.

		The presence of both, systematic acceleration (first-order Fermi), when particles cross a reconnecting CS, and stochastic heating (second-order Fermi), when particles interact with non-reconnecting CSs (turbulence) is apparent. 	
		Non-reconnecting CSs (and thus stochastic acceleration) dominate the heating. As a result, their role in the formation of  non-thermal distributions is minimal. On the other hand, reconnecting CSs mostly contribute to the acceleration of particles and the formation of  power-law tails, and they have a minimal contribution to the heating of a plasma. These results are analyzed in detail in our recent article \cite{Vlahos21}.
		 
		 In summary, we can conclude the following:
		 
		 \begin{enumerate}
		 	\item Coronal heating is not due to magnetic reconnection and nano-flares. It is rather due to non-reconnecting CSs formed by the mechanism suggested by \cite{Parker83}. On the other hand, reconnecting current sheets will mostly contribute to the acceleration of particles. Thus, the presence or reconnecting current sheets in a system becomes apparent by the formation of power-law tails in the energy distributions.
		 	\item The recently discovered "campfires" \citep{ChenB2021} may be a step towards the direction of observing the coexistence of non-reconnecting and reconnecting CSs in the solar corona. 
		 	\item In confined and explosive solar flares, the impulsive heating and acceleration of particles are manifest. Thus, the presence of both, turbulent heating and magnetic reconnection, is obvious from current observations. 
		 	\item In the solar wind the co-existence of non-reconnecting and reconnecting CSs, which contribute to the stochastic heating and  the formation of the non-thermal tail, respectively, is also well established.  We expect that works investigating Parker Solar Probe data will shed light on this point in the near future.
		 	\item The Earth's magnetotail and the magnetopause are also an example of the coexistence of reconnecting and non-reconnecting CSs, and the heating and acceleration of particles is once again obvious in the data analyzed so far. 
		 	\item In the downstream environment of shocks, the coupling of turbulent non-reconnecting CSs and reconnecting CSs is present and contributes to the heating and acceleration of  particles.
		 	\item In astrophysical jets, the co-existence of turbulent stochastic heating and particle acceleration through systematic acceleration (reconnecting CSs or shocks) is present in the strongly turbulent plasma flows  \citep{Manolakou99}.
		 \end{enumerate}
		 
		  These are only a few examples of astrophysical and space plasmas that illustrate  the intensive coupling of turbulence (non-reconnecting CSs), which dominates the stochastic heating, and magnetic reconnection (reconnecting CSs), which dominates the formation of power-law tails in energy distributions.
		 
		  \vspace{0.2 cm}
		 
		Let us stress that it is now a well-accepted fact that magnetic reconnection and turbulence are going hand in hand in several astrophysical  environments. What we added to this well-accepted fact with our current work in \cite{Vlahos21} is that stochastic heating by turbulence (non-reconnecting CSs) and systematic acceleration by reconnecting CSs are responsible for the impulsive heating and the acceleration of particles, respectively. We suggest that depending on the large scale magnetic topology, which is responsible for setting up the turbulent reconnection environment, we may have several combinations of heating and acceleration of particles, e.g.: 
		 
		 \begin{itemize}
		 	\item Steady or impulsive heating (soft X-Ray dominated flares),  when non-reconnecting CSs (turbulence) dominate \citep{Einaudi21, Hudson21}.
		 	\item Impulsive heating and particle acceleration when both, non-reconnecting and reconnecting CSs, are present \citep{Lin03, ChenB2021}.
		 	\item Acceleration of high energy particles and no impulsive heating ("cold" flares) when reconnecting current sheets dominate \citep{Lysenko_2018}.
		 \end{itemize}
		 
		 There are two approaches active today in the study of magnetic energy dissipation in plasmas. The one uses isolated reconnecting CSs as the basic model, and the other one turbulent reconnection, where non-reconnecting CSs and turbulence co-exist with reconnecting CSs. If we choose the first, we have troubles interpreting the steady or impulsive heating of plasmas. If we choose the second, heating and acceleration of high-energy particles are naturally explained.
		 	
		 	  \vspace{0.2 cm}

		  After 50 years of active research, the coronal heating problem remains an open problem. Currently, theoretical models proposed to explain the heating of the corona are split into two categories: waves and reconnecting CSs (nano-flares). The purpose of this comment is that of suggesting an alternative heating mechanism, namely non-reconnecting CSs formed by the turbulent convection zone in complex magnetic topologies.

		%\bibliographystyle{aa}
%\bibliography{vlahosturb2}

\end{document}